\documentclass[runningheads]{llncs}
\usepackage[T1]{fontenc}
\usepackage{graphicx}
\usepackage[nolist]{acronym}
\usepackage{float}
\usepackage{fontawesome}
\usepackage{graphicx}
\usepackage{amsmath}
\usepackage{tikz}
\usepackage[ruled,vlined]{algorithm2e}
\usepackage{graphicx}
\usepackage[table]{xcolor}
\usepackage{bm}
\usepackage{booktabs} 
\usepackage{amssymb}
\usepackage{amsfonts}

\usepackage{graphicx}

\usepackage{url}
\begin{document}
%Numbering
\acrodef{3GPP}{3rd Generation Partnership Project}
\acrodef{5G}{5th Generation Mobile Network}
\acrodef{6G}{6th Generation Mobile Network}
%-----A-----
\acrodef{AI}{Artificial Intelligence}
\acrodef{AI4Net}{\ac{AI} for Networking}
\acrodef{MDP}{Markov Decision Process}
\acrodef{AIDER}{Aerial Image Dataset for Emergency Response}
\acrodef{AMF}{Access and Mobility Management Function}
\acrodef{AIaaS}{Artificial Intelligence-as-a-Service}
\acrodef{AC}{Actor-Critic}
\acrodef{IID}{Independent and Identically Distributed}
%-----B-----
\acrodef{B5G}{Beyond Fifth Generation}
\acrodef{BPF}{Berkeley Packet Filter}
%-----C-----
\acrodef{CBR}{Constant Bit Rate}
\acrodef{CSV}{Comma-Separated Values}
\acrodef{CPU}{Central Processing Unit}
\acrodef{CNN}{Convolutional Neural Network}
\acrodef{CNNs}{Convolutional Neural Networks}
\acrodef{C-V2X}{Cellular Vehicle to-Everything}
%-----D-----
\acrodef{DoS}{Denial of Service}
\acrodef{DDQL}{Double Deep
 Q-learning}
\acrodef{DDoS}{Distributed Denial of Service}
\acrodef{DDPG}{Deep Deterministic Policy Gradient}
\acrodef{DNN}{Deep Neural Network}
\acrodef{DRL}{Deep Reinforcement Learning}
\acrodef{DQN}{Deep Q-Network}
\acrodef{DT}{Decision Tree}
\acrodef{DDQN}{Double Deep Q-Network}
\acrodef{DN}{Data Network}
%-----E-----
\acrodef{ETSI}{European Telecommunications Standards Institute}
\acrodef{eNWDAF}{Evolved Network Data Analytics Function}
\acrodef{eBPF}{Extended Berkeley Packet Filter}
%-----F-----
\acrodef{FIBRE}{Future Internet Brazilian Environment for Experimentation}
%-----G-----
\acrodef{GNN}{Graph Neural Networks}
\acrodef{GPU}{Graphics Processing Unit}
\acrodef{GTP}{GPRS Tunnelling Protocol}
%-----H-----
\acrodef{HTM}{Hierarchical Temporal Memory}

%-----I-----
\acrodef{IAM}{Identity And Access Management}
\acrodef{ICMP}{Internet Control Message Protocol}
\acrodef{IID}{Informally, Identically Distributed}
\acrodef{IoE}{Internet of Everything}
\acrodef{IoT}{Internet of Things}
\acrodef{ITU}{International Telecommunication Union}
\acrodef{ISP}{Internet Service Provider}
\acrodef{ISPs}{Internet Service Providers}
%-----J-----
%-----K-----
\acrodef{KNN}{K-Nearest Neighbors}
\acrodef{KPI}{Key Performance Indicator}
\acrodef{KPIs}{Key Performance Indicators}
%-----L-----
\acrodef{LSTM}{Long Short-Term Memory}
%-----M-----
\acrodef{MAE}{Mean Absolute Error}
\acrodef{ML}{Machine Learning}
\acrodef{MLaaS}{Machine Learning as a Service}
\acrodef{MOS}{Mean Opinion Score}
\acrodef{MAPE}{Mean Absolute Percentage Error}
\acrodef{MSE}{Mean Squared Error}
\acrodef{MEC}{Multi-access Edge Computing}
\acrodef{mMTC}{Massive Machine Type Communications}
\acrodef{MFA}{Multi-factor Authentication}
\acrodef{MLP}{Multi-Layer Perceptron}
\acrodef{MADRL}{Multi-Agent Deep Reinforcement Learning}
\acrodef{MAB}{Multi-Armed Bandit}
\acrodef{MILP}{Mixed Integer Linear Programming}
%-----N-----
\acrodef{NWDAF}{Network Data Analytics Function}
\acrodef{Net4AI}{Networking for \ac{AI}}
\acrodef{NS}{Network Slicing}
\acrodef{NFV}{Network Function Virtualization}
%-----O-----
\acrodef{OSM}{Open Source MANO}
%-----P-----
\acrodef{PCA}{Principal Component Analysis}
\acrodef{PoC}{Proof of Concept}
\acrodef{PPO}{Proximal Policy Optimization}
\acrodef{POMDP}{Partially Observable Markov decision process}
%-----Q-----
\acrodef{QoE}{Quality of experience}
\acrodef{QoS}{Quality of Service}
%-----R-----
\acrodef{RAM}{Random Access Memory}
\acrodef{RF}{Random Forest}
\acrodef{RL}{Reinforcement Learning}
\acrodef{RMSE}{Root Mean Square Error}
\acrodef{RNN}{Recurrent Neural Network}
\acrodef{RTT}{Round-Trip Time}
\acrodef{RAN}{Radio Access Network}

%-----S-----
\acrodef{SDN}{Software-Defined Networking}
\acrodef{SFI2}{Slicing Future Internet Infrastructures}
\acrodef{SLA}{Service-Level Agreement}
\acrodef{SON}{Self-Organizing Network}
\acrodef{SMF}{Session Management Function}
\acrodef{S-NSSAI}{Single Network Slice Selection Assistance Information}
\acrodef{SVM}{Support Vector Machine}
\acrodef{SOPS}{Service-Aware Optimal
 Path Selection}
 \acrodef{SRv6}{Segment Routing over IPv6}
%-----T-----
\acrodef{TQFL}{Trust Deep Q-learning Federated Learning}
\acrodef{TEID}{Tunnel Endpoint Identifier}
\acrodef{TEIDs}{Tunnel Endpoint Identifiers}
%-----U-----
\acrodef{UE}{User Equipment}
\acrodef{UEs}{User Equipments}
\acrodef{UPF}{User Plane Function}
\acrodef{UPFs}{User Plane Functions}
\acrodef{PDU}{Packet Data Unit}
\acrodef{URLLC}{Ultra-Reliable and Low Latency Communications}
\acrodef{UAV}{Unmanned Aerial Vehicle}
\acrodef{UAVs}{Unmanned Aerial Vehicles}
%-----V-----
\acrodef{VoD}{Video on Demand}
\acrodef{VR}{Virtual Reality}
\acrodef{AR}{Augmented Reality}
\acrodef{V2V}{Vehicle-to-Vechile}
\acrodef{V2X}{Vehicle-to-Everything}
\acrodef{VNF}{Virtual Network Function}
\acrodef{VNFs}{Virtual Network Functions}

%-----W-----
%-----X-----
\acrodef{XDP}{eXpress Data Path}
\acrodef{XR}{eXtended Reality}
%-----Y-----
%-----Z-----
%
\title{An Intelligent eUPF for Time-Sensitive Path Selection in B5G Edge Networks}
%
%\titlerunning{Passive Latency to Intelligent UPF Path Selection}
%
\author{
Rodrigo Moreira\inst{1} \and
Larissa F. Rodrigues Moreira\inst{1} \and
Tereza C. Carvalho\inst{2} \and
Flávio de Oliveira Silva\inst{3}
}
\authorrunning{R. Moreira et al.}
\institute{
Federal University of Viçosa, Minas Gerais, Brazil \\
\email{\{rodrigo,larissa.f.rodrigues\}@ufv.br}
\and
University of São Paulo, São Paulo, Brazil \\
\email{terezacarvalho@usp.br}
\and
University of Minho, Braga, Portugal \\
\email{flavio@di.uminho.pt}
}
\maketitle              % typeset the header of the contribution
\begin{abstract}
In Beyond 5G (B5G) networks, intelligent, flexible traffic management is essential to meet the stringent speed and reliability requirements of new applications. This paper presents an improved User Plane Function (eUPF) design that uses a Deep Q-Network (DQN) agent for real-time path selection between Multi-access Edge Computing (MEC) and cloud endpoints. The path selection problem is formulated as a Partially Observable Markov Decision Process (POMDP). We propose a novel passive delay measurement method that uses eBPF programs to link TEID-based timestamps in GTP-U traffic, allowing for low-cost delay estimation without active testing. Experiments show that the DQN agent substantially outperforms a random baseline, with lower average latency, more stable rewards, and more reliable low-delay path choices. These results demonstrate the effectiveness of AI-driven control in B5G core networks and the promise of reinforcement learning for modern network management.

\keywords{Beyond 5G Networks \and User Plane Function \and Reinforcement Learning \and eBPF/XDP \and Multi-access Edge Computing \and Time-Sensitive Path Selection}
\end{abstract}
\section{Introduction}\label{sec:introduction}

The increasing prevalence of compute-intensive applications, coupled with the diversity of mobile devices, results in substantial data volumes in \ac{5G}. To accommodate this demand, especially in the \ac{B5G} context, it is crucial to minimize service access delay by bringing services closer to users through approaches such as \ac{MEC}. Industries such as automation, autonomous driving, and cloud gaming, for instance, impose stringent ultralow-latency requirements~\cite{silva2025}, and in response, placing network functions near the edge not only helps meet these demands but also offers a promising way to alleviate backbone traffic surges, particularly for \ac{NFV}-based components~\cite{Bellin2024}. This drives ongoing research on efficient path-selection strategies in the data plane of mobile networks.

Building on these advancements, the state of the art presents various approaches for service steering and resource allocation within the broader context of path selection~\cite{moreira_nasor}, particularly across two key segments of mobile networks. The first research stream focuses on the \ac{RAN}, where strategies range from traffic-aware \ac{RAN} slicing to open and disaggregated solutions, such as Open \ac{RAN}~\cite{Agarwal2025}. The second stream focuses on traffic steering in the mobile network data plane, with particular emphasis on solutions to enhance the performance of the \ac{UPF} \cite{Tran2024}. Despite this progress, the literature identifies further opportunities to explore how the \ac{UPF} can be instrumented to enable greater programmability and intelligent behavior~\cite{Chen2025}.

Recent efforts to enhance programmability and intelligent behavior in the mobile core have led to the introduction of analytics capabilities in the \ac{3GPP} Release 16, facilitated by the \ac{NWDAF} entity~\cite{Garcia-Martin2024}. However, the mechanisms for data collection and component instrumentation have posed challenges across various deployments~\cite{Larissaw6g}.
This study explores the potential of transforming the \ac{UPF} into a more intelligent entity, capable of considering latency by examining \ac{GTP} \ac{TEIDs} and employing a \ac{RL} mechanism to optimize its real-time performance.

In this work, we introduce and assess the \ac{UPF} instrumentation for estimating the end-user latency of the \ac{UE} by analyzing the time-shifting of \ac{TEIDs} arriving at the \ac{UPF}. This approach empowers the \ac{UPF} to autonomously select the optimal interface for traffic forwarding and apply intelligent traffic steering based on latency awareness, thereby reducing end-to-end delay. The efficacy of our methodology is demonstrated through a \ac{MEC} and cloud-based application, validating how the \ac{UPF}, in conjunction with an \ac{RL} agent, can non-intrusively direct traffic along various paths by monitoring \ac{UE} latency and using \ac{XDP}.  This design positions the e\ac{UPF} as an intelligent user-plane element for time-sensitive path selection across \ac{MEC} and cloud environments.

The contributions of this study are as follows: (i) the development of a novel \ac{UPF} instrumentation for \ac{UE} latency estimation; (ii) a comparative evaluation of a naive \ac{UPF} path selection versus an \ac{RL}-based approach; (iii) the design and implementation of a non-intrusive telemetry mechanism using \ac{XDP} to enable real-time TEID monitoring and latency estimation at the \ac{UPF}; and (iv) an end-to-end experimental validation showcasing how the proposed architecture improves latency-sensitive service delivery in \ac{MEC} and cloud scenarios.

The remainder of this paper is organized as follows: Section~\ref{sec:related_work} presents representative approaches for \ac{MEC}-aware decision-making in 5G/6G. In Section~\ref{sec:proposed_approach}, we present our method for intelligent intervention in the dataplane. In Section~\ref{sec:evaluation_setup}, we describe the technologies used to perform the proof of concept. In Section~\ref{sec:results_and_discussion}, we present the discussions and insights gained from our approach. Finally, in Section~\ref{sec:concluding_remarks}, we provide the conclusions and lessons learned.

\section{Related Work}\label{sec:related_work}

\begin{table}[!ht]
\centering
\caption{Summary of representative approaches for \ac{MEC} aware decision making in Fifth Generation and Sixth Generation networks.}
\label{tab:sota_compact_yesno}
\setlength{\tabcolsep}{4pt}
\renewcommand{\arraystretch}{1.05}
\resizebox{\columnwidth}{!}{%
\begin{tabular}{lccccc}
\hline
\rowcolor[HTML]{EFEFEF}
\textbf{Paper} & \textbf{Learning} & \textbf{Optimization} & \textbf{User Plane Function} & \textbf{Tunnel Endpoint Identifier} & \textbf{Real Testbed} \\
\hline
Shokrnezhad et al.~\cite{Shokrnezhad2024} &  \faCircle &  \faCircle &  \faCircleO  &  \faCircleO  &  \faCircleO \\
Tran et al.~\cite{Tran2024}              &  \faCircleO  &  \faCircle &  \faCircle &  \faCircleO  &  \faCircle \\
Gan et al.~\cite{Gan2022}                &  \faCircle &  \faCircle &  \faCircleO  &  \faCircleO  &  \faCircleO \\
Maleki et al.~\cite{Maleki2024}          &  \faCircle &  \faCircle &  \faCircleO  &  \faCircleO  &  \faCircleO \\
Sasithong et al.~\cite{Sasithong2025}    &  \faCircle &  \faCircle &  \faCircle &  \faCircleO  &  \faCircleO \\
Hisyam et al.~\cite{Hisyam2024}          &  \faCircle &  \faCircle &  \faCircleO  &  \faCircleO  &  \faCircleO \\
Nguyen et al.~\cite{Nguyen2021}          &  \faCircle &  \faCircle &  \faCircle &  \faCircleO  &  \faCircleO \\
Pimpalkar et al.~\cite{Pimpalkar2025}    &  \faCircleO  &  \faCircle &  \faCircleO  &  \faCircleO  &  \faCircleO \\
Kibalya et al.~\cite{Kibalya2025}        &  \faCircle &  \faCircle &  \faCircle &  \faCircleO  &  \faCircleO \\
Ibrahimi et al.~\cite{Ibrahimi2024}      &  \faCircle &  \faCircle &  \faCircleO  &  \faCircleO  &  \faCircleO \\
\hline
\textbf{Our Approach}                    & \faCircle & \faCircle & \faCircle & \faCircle & \faCircle \\
\hline
\end{tabular}%
}
\end{table}

Automated management systems are essential for emerging \ac{XR} and the metaverse \cite{Donatti2024}, \cite{Christopoulou2025}. Recent research focus includes instrumenting different segments of \ac{5G}, including the \ac{RAN}, core, and transport, to improve adaptability and performance monitoring. This section reviews studies that use the \ac{UPF} to measure \ac{UE} connectivity parameters, adapt to dynamic traffic conditions, and enable \ac{ISPs} to track and enforce \ac{SLA} guarantees effectively.

Deep Reinforcement Learning is widely applied for intelligent resource management. Shokrnezhad et al. \cite{Shokrnezhad2024} address joint \ac{VNF} placement, traffic prioritization, and path selection in \ac{B5G} networks under latency and capacity constraints using \ac{DDQL} Double Deep Q learning. Gran et al. \cite{Gan2022} use \ac{MADRL} with collaborative \ac{DQN} models for computation offloading in \ac{5G} \ac{MEC} systems to minimize total system cost. 

Similarly, Sasithong et al. \cite{Sasithong2025} apply \ac{DQN} and \ac{AC} algorithms for dynamic \ac{UPF} allocation in \ac{C-V2X} \cite{Sasithong2025}. Nguyen et al. \cite{Nguyen2021} utilize \ac{PPO} and \ac{SVM} for scaling \ac{UPF} instances \cite{Nguyen2021}, while Kibalya et al. \cite{Kibalya2025} employ policy networks for joint \ac{UPF} and application placement in multi-slice \ac{MEC} \cite{Kibalya2025}. For \ac{URLLC}, Ibrahimi et al. \cite{Ibrahimi2024} implement a \ac{DQN} strategy to balance congestion and throughput \cite{Ibrahimi2024}.

Other research employs heuristics, optimization, or supervised learning. Tran et al. \cite{Tran2024} introduce a rule-based computing-aware traffic steering architecture for \ac{5G} to route traffic to optimal \ac{MEC} instances using \ac{SRv6} \cite{Tran2024}. Maleki et al. \cite{Maleki2024} utilize \ac{MAB} and \ac{DNN} models for \ac{QoS} aware content delivery under mobility \cite{Maleki2024}. In \ac{6G} contexts, Pimpalkar et al. \cite{Pimpalkar2025} formulate a multi-constraint convex problem for \ac{SOPS} \cite{Pimpalkar2025}. Hisyam Ng and Mahmoodi \cite{Hisyam2024} propose a supervised learning scheme using Random Forests and \ac{MILP} for dynamic steering \cite{Hisyam2024}.

Table~\ref{tab:sota_compact_yesno} summarizes representative approaches for \ac{MEC} aware decision making in Fifth Generation and Sixth Generation networks using a binary comparison of key capabilities, namely whether each work employs learning, performs explicit optimization, targets the User Plane Function, incorporates Tunnel Endpoint Identifier awareness, and validates results on a real testbed. The table shows that prior studies largely rely on simulation and do not exploit Tunnel Endpoint Identifier signals for data-plane decisions, even when learning or optimization is applied. In contrast, our approach integrates a \ac{DQN}-based reinforcement learning agent with User Plane Function-level control and explicit Tunnel Endpoint Identifier awareness, and it is validated on a real testbed, enabling passive latency-driven N6 interface selection for Beyond Fifth Generation scenarios spanning \ac{MEC} and cloud paths.

\section{Proposed Approach}\label{sec:proposed_approach}

\begin{figure}[!htbp]
\centering
  \includegraphics[width=0.7\columnwidth]{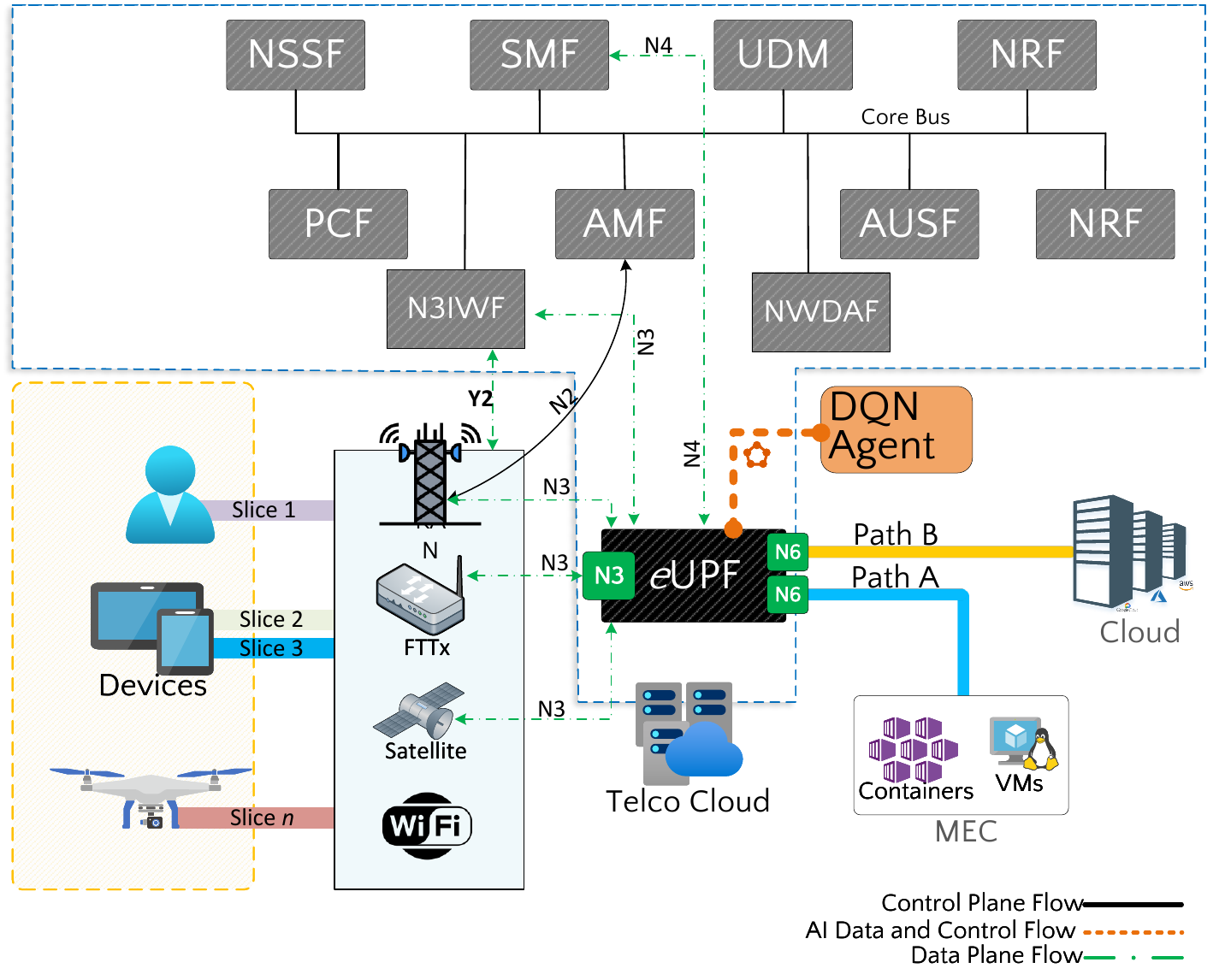}
  \caption{Proposed Approach.}
  \label{fig:method}
\end{figure}

Building on the insights and limitations identified in prior work, we propose an enhancement to the traditional \ac{UPF}, hereafter referred to as e\ac{UPF} (enhanced \ac{UPF}), which integrates an \ac{RL} agent to enable intelligent, real-time path selection for individual network slices. As shown in Fig.~\ref{fig:method}, the \ac{DQN}-based agent observes latency and network conditions to dynamically steer traffic across multiple N6 paths (e.g., Path~A and Path~B), targeting either \ac{MEC} or Cloud endpoints. This architecture enables the \ac{UPF} to surpass static policies by continuously adapting to network dynamics, user mobility, and service-level requirements. By leveraging AI control at the data plane, the proposed e\ac{UPF} architecture enables smarter, slice-aware forwarding decisions aligned with the vision of agile and autonomous 5G/6G core networks.

\subsection{Problem Formulation}

We formulate adaptive path selection as a Partially Observable Markov Decision Process, where the enhanced \ac{UPF} applies a slice specific forwarding policy at run time, and the reinforcement learning agent makes per flow decisions using the \ac{TEID} as the in kernel key that identifies the active \ac{PDU} session associated with that slice context.

\textbf{States and Transitions.} Let $S_t = (s_{n6a}(t), s_{n6b}(t))$ denote the latent environment state at time $t$, where $s_i(t) \in \{\text{GOOD}, \text{BAD}\}$ for $i \in \{\texttt{n6a}, \texttt{n6b}\}$. Transitions from GOOD to BAD occur independently with fixed probabilities $p_i$ per packet. Once degraded, each interface remains in the \text{BAD} state for a deterministic duration $D_i$ before reverting to the \text{GOOD} state. These semi Markov dynamics introduce memory that must be inferred from observations.

\textbf{Observations.} The agent does not observe $S_t$ directly. Instead, after selecting an action $A_t \in \{\texttt{n6a}, \texttt{n6b}\}$, it receives a noisy delay measurement. Let $i$ denote the interface selected by $A_t$ (that is, $i = A_t$). The observation process (Eq.~1) is:

{\footnotesize \begin{equation} \begin{aligned} RTT(A_t) &= d_i \cdot \mathbb{I}[s_i(t) = \text{BAD}] + \xi_t \\ \xi_t &\sim \mathcal{U}(-J_{\max}, +J_{\max}) \label{eq:jitter_model} \end{aligned} \end{equation} }

where $d_i$ is the additional delay incurred when interface $i$ is in the BAD state, $\mathbb{I}[\cdot]$ is the indicator function, equal to 1 if the condition holds and 0 otherwise, and $\xi_t$ models bounded measurement noise (jitter) with maximum amplitude $J_{\max}$.

Because the agent operates under partial observability, we define a scalar observed state $s_t$ derived from the latest measurement. In our implementation, $s_t$ corresponds to the most recent per \ac{TEID} delay proxy exposed by the data plane. This $s_t$ is the input used by the learning agent at decision time $t$.

\textbf{Actions and Rewards.} At each time step $t$, the agent selects an interface $A_t \in \{\texttt{n6a}, \texttt{n6b}\}$ via a policy $\pi(a_t|s_t)$. We define the reward as the negative observed delay (Eq.~\ref{eq:reward}):

\begin{equation}
R_t = -RTT(A_t).
\label{eq:reward}
\end{equation}

We normalize $R_t$ via min max scaling to stabilize training, mapping rewards to a bounded range while preserving ordering, so that lower delays correspond to higher normalized rewards.

\textbf{Learning Objective.} The agent aims to maximize expected discounted return (Eq.~\ref{eq:objective}):

\begin{equation}
\max_{\pi}\; \mathbb{E}\!\left[\sum_{t=0}^{T}\gamma^{t}R_{t}\right],
\label{eq:objective}
\end{equation}

where $\gamma \in (0, 1]$ is the discount factor, $T$ is the episode horizon in decision steps, and the expectation is taken over trajectories induced by policy $\pi$ and the environment dynamics. Because $S_t$ is latent and only $s_t$ is observed, deep reinforcement learning, particularly \ac{DQN}, is suitable to learn effective decisions from sequences of noisy measurements.

\subsection{\ac{TEID}-Correlated Delay Measurement}

To avoid injecting additional active probes, we leverage a passive data plane measurement approach based on \ac{TEID} correlated timestamps within the \ac{UPF}. The system attaches eBPF programs directly to the \ac{GTP}-U traffic path, extracting the \ac{TEID} from each packet. For each \ac{TEID} $i$, a stateful map stores a request timestamp $t_i^{\text{req}}$ when a request packet is observed.

When the corresponding response packet for the same \ac{TEID} $i$ is observed at the \ac{UPF}, the program computes an in network round trip proxy (Eq.~\ref{eq:teid_proxy}):

\begin{equation}
\Delta t_i = t_i^{\text{resp}} - t_i^{\text{req}}.
\label{eq:teid_proxy}
\end{equation}

Here, $t_i^{\text{req}}$ is the timestamp captured for the request, and $t_i^{\text{resp}}$ is the timestamp captured for the matched response, both observed at the \ac{UPF} for the same \ac{TEID} $i$. In our proof of concept, the request response pattern is instantiated with \ac{ICMP} echo traffic, and the response is identified by matching the echo reply to the previously observed echo request within the same \ac{TEID} context, which provides a consistent pairing signal for timestamp correlation.

This measurement is a \ac{UPF} observed delay proxy, not a fully instrumented end to end RTT across the entire \ac{DN} and \ac{RAN}. Its purpose is to provide a low overhead, online signal that reflects the experienced user plane delay under the current forwarding choice, enabling the agent to adapt decisions even when only partial visibility is available.

To implement the passive latency measurement, we developed an in kernel routine that associates \ac{TEID} identified flows with timestamped events at the \ac{UPF} data plane. Algorithm~\ref{alg:teid_latency_measurement} presents the logic executed by the eBPF program attached to the \ac{GTP}-U traffic path. Upon receiving a packet, the program extracts the corresponding \ac{TEID} and records a request timestamp if none already exists. When a subsequent packet for the same \ac{TEID} arrives, it computes the delay proxy and resets tracking state, enabling continuous measurement without additional probe injection.

\begin{algorithm}[!htbp]
\caption{TEID-Correlated \ac{RTT} Measurement.}
\scriptsize
Initialize \texttt{rtt\_map} as a hash map with key: \ac{TEID} and value: \texttt{round\_trip} structure\;
\SetKwFunction{FMain}{measure\_rtt}
\SetKwProg{Fn}{Function}{:}{}
\Fn{\FMain{skb}}{
    Extract \ac{TEID} from incoming \ac{GTP}-U packet\;
    $now \gets$ current kernel timestamp (\texttt{bpf\_ktime\_get\_ns()})\;
    Lookup entry in \texttt{rtt\_map} by \ac{TEID}\;
    \eIf{entry does not exist}{
        Insert new entry with \texttt{ts\_request} $\gets now$\;
    }{
        \eIf{\texttt{ts\_request} is zero}{
            Set \texttt{ts\_request} $\gets now$\;
        }{
            Compute RTT: \texttt{last\_rtt} $\gets now - \texttt{ts\_request}$\;
            Reset \texttt{ts\_request} $\gets 0$\;
            Increment \texttt{count}\;
        }
    }
}
\label{alg:teid_latency_measurement}
\end{algorithm}

\section{Evaluation Setup}\label{sec:evaluation_setup}

\begin{figure}[htbp]
\centering
  \includegraphics[width=0.7\columnwidth]{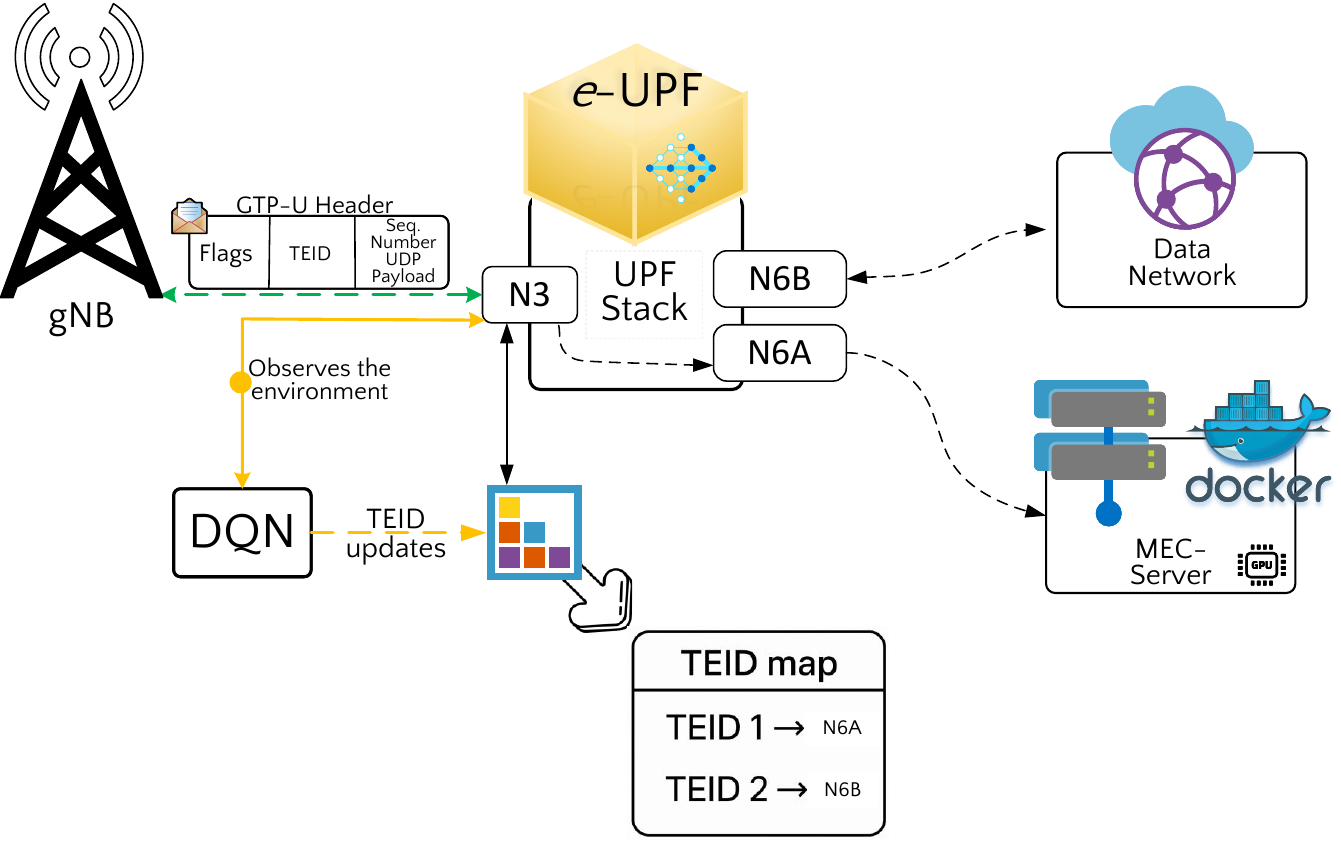}
  \caption{Experimental Setup.}
  \label{fig:experimental_setup}
\end{figure}

Our evaluation was conducted on the FABRIC testbed, using a high performance virtual machine provisioned with 32\,GB of RAM and 32 vCPUs. The environment was orchestrated with Kubernetes v1.28.15 running on Ubuntu, and Free5GC was used to emulate the 5G core network. Fig.~\ref{fig:experimental_setup} illustrates the deployment, where the \ac{DQN} agent interacts with the user plane through \ac{eBPF}/\ac{XDP} hooks and shared maps. We leveraged a custom BPF toolset to implement programmable datapath control, enabling dynamic action updates and observation retrieval directly at the \ac{UPF}. This setup supports reproducible experimentation with low level latency estimation and adaptive path selection across \texttt{n6a} and \texttt{n6b} interfaces.

The stochastic degradation parameters used in our experiments are summarized in Table~\ref{tab:experimental_setup_params}. For each interface $i \in \{\texttt{n6a}, \texttt{n6b}\}$, a degradation event is triggered independently at each packet arrival according to a failure probability $p_i$. Once triggered, the interface remains degraded for a fixed failure duration $D_i$ before automatically recovering. When an interface is in the degraded state, an additional delay $d_i$ is induced, matching the observation model in Eq.~\ref{eq:jitter_model}. A bounded jitter term with maximum amplitude $J_{\max}$ is applied globally across both paths, as defined in Eq.~\ref{eq:jitter_model}.

\begin{table}[ht]
\centering
\caption{Network Path Parameters for Delay Induction Scenario.}
\label{tab:experimental_setup_params}
\resizebox{\columnwidth}{!}{%
\begin{tabular}{lccc}
\toprule
\textbf{Path} & \textbf{Failure Probability} & \textbf{Failure Duration (s)} & \textbf{Bad State Delay (ms)} \\
\midrule
Path A (\ac{MEC}) & 0.01 & 10 & 800 \\
Path B (Cloud) & 0.10 & 20 & 800 \\
\midrule
\multicolumn{4}{l}{\textbf{Global Parameter:} Maximum Jitter $J_{\max}$ = 3 ms} \\
\bottomrule
\end{tabular}
}
\end{table}

\subsection{DQN Based Path Selection Architecture}\label{subsec:dqn_based_architecture}

The learning component is implemented as a \ac{DQN}, which approximates the action value function $Q(s_t, a_t; \theta)$ with a deep neural network parameterized by $\theta$. The agent receives as input a scalar state $s_t$ derived from the most recent per \ac{TEID} delay proxy, and outputs Q values for the two discrete actions corresponding to the available egress interfaces, namely $\texttt{n6a}$ and $\texttt{n6b}$. Action selection follows an $\epsilon$ greedy exploration strategy, where $\epsilon$ decays from an initial exploration probability to a smaller final value over training.

The network architecture is a feedforward multilayer perceptron with two hidden layers of 64 units each and ReLU activations. Learning is stabilized using experience replay and a separate target network with parameters $\theta^{-}$ that are periodically synchronized from the online network. At each decision step, the agent stores a transition $(s_t, a_t, r_t, s_{t+1})$ in a replay buffer and samples mini batches to update $\theta$ by minimizing the temporal difference objective defined in Eq.~5. Table~\ref{tab:dqn_params} summarizes the hyperparameters and model configuration used in all experiments.

\begin{table}[!htb]
\centering
\caption{Hyperparameters and Network Architecture of the \ac{DQN} Agent.}
\label{tab:dqn_params}
\begin{tabular}{cc}
\toprule
\textbf{Parameter} & \textbf{Value} \\
\midrule
Discount factor ($\gamma$)           & 0.99 \\
Learning rate (Adam) ($\alpha$)      & $5 \times 10^{-4}$ \\
Batch size                           & 32 \\
Replay memory size                   & 2000 transitions \\
Exploration start ($\epsilon_{\text{start}}$) & 0.9 \\
Exploration end ($\epsilon_{\text{end}}$)     & 0.01 \\
Exploration decay rate               & 0.990 \\
Target network update frequency      & Every 5 episodes \\
Training episodes                    & 400 \\
\midrule
\textbf{Network Architecture} & \\
Input dimension                 & 1 (scalar state) \\
Hidden layers                  & 2 × 64 units (Linear + ReLU) \\
Output dimension               & 2 (Q values for \texttt{n6a}, \texttt{n6b}) \\
\bottomrule
\end{tabular}
%\label{tab:parameters_and_dqn_architecture}
\end{table}

\subsection{Agent Environment Interaction via XDP Maps}\label{subsec:xdp_agent_interaction}

The interaction between the \ac{DQN} agent and the forwarding plane is implemented using two \ac{XDP} maps that serve as the interface between learning logic and packet processing. The first map exposes observations to the agent by providing per \ac{TEID} delay values computed as the \ac{UPF} observed proxy $\Delta t_i$ in Eq.~\ref{eq:teid_proxy}, obtained by correlating \ac{TEID} keyed timestamps for request and response packets observed at the same \ac{UPF} datapath. In the proof of concept, \ac{ICMP} traffic provides the request response pattern required for pairing, while the measurement itself remains passive with respect to additional monitoring probes.

The second map enables actions by storing the selected egress interface per \ac{TEID}, so that packet forwarding for that \ac{PDU} session follows the action chosen by the agent. Specifically, at each decision time $t$ the agent selects $a_t \in \{\texttt{n6a}, \texttt{n6b}\}$ and writes the corresponding value into the action map, which the \ac{XDP} program reads to enforce forwarding on the chosen N6 path.

Learning follows the standard \ac{DQN} update rule, minimizing the temporal difference loss, as Eq.~5:

\begin{align}
\mathcal{L}(\theta) &=
\mathbb{E}_{(s_t,a_t,r_t,s_{t+1})}\!\Big[
\big(r_t + \gamma \max_{a'} Q(s_{t+1}, a'; \theta^-)
\\ \nonumber
&\qquad - Q(s_t, a_t; \theta)\big)^2
\Big]
\label{eq:dqn_loss}
\end{align}

where $Q(s_t, a_t; \theta)$ is the online action value network, $\theta^-$ denotes the target network parameters, $\gamma$ is the discount factor (Table~\ref{tab:dqn_params}), and $(s_t, a_t, r_t, s_{t+1})$ are transitions sampled from the replay buffer. The reward $r_t$ is computed from the observed delay using Eq.~\ref{eq:reward} and then normalized by min max scaling, ensuring stable gradients during training. Episodes last 60 seconds, during which the agent repeatedly reads the latest per \ac{TEID} observation state $s_t$, selects an action, updates the action map, and stores the resulting transition for learning.

This design enables the \ac{DQN} agent to translate passive latency observations into fine grained forwarding control, achieving adaptive N6 interface selection under stochastic path degradations.

\section{Results and Discussion}\label{sec:results_and_discussion}

Building on the experimental setup and the agent environment interaction described above, this section analyzes learning dynamics, policy behavior, and latency outcomes achieved by the proposed \ac{DQN} based path selection strategy. We compare it against a random baseline across multiple metrics, including reward evolution, interface selection patterns, and measured round trip time.

\begin{figure}[!ht]
    \centering
    \begin{tabular}{@{}c@{}c@{}}
        \includegraphics[width=0.450\columnwidth]{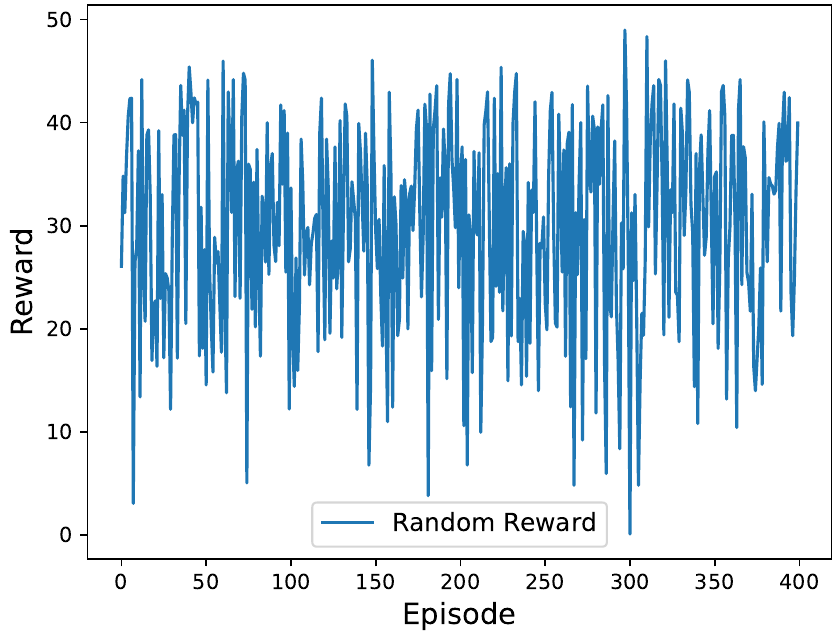} &
        \includegraphics[width=0.450\columnwidth]{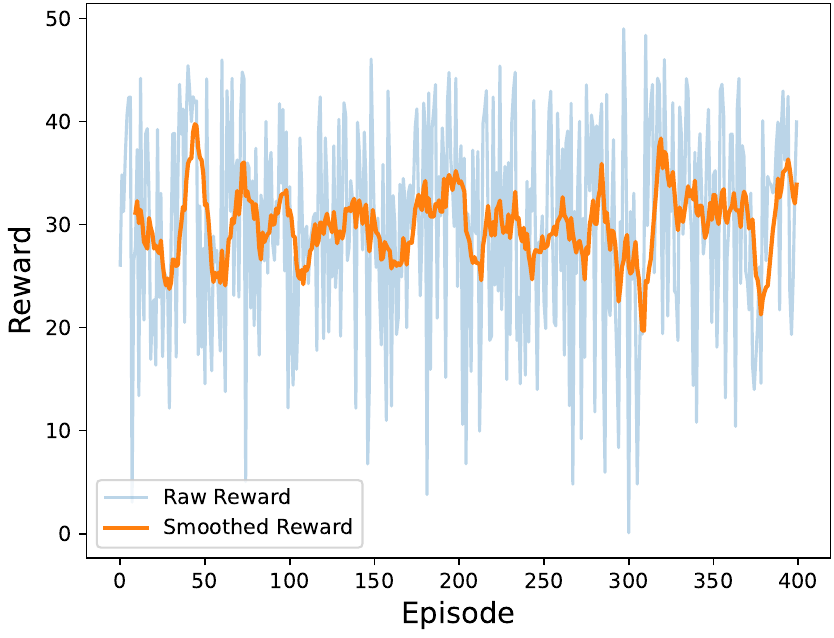} \\
        \small (a) Raw reward per episode. & \small (b) Smoothed reward.  \\
    \end{tabular}
    \caption{Comparison of raw and smoothed cumulative rewards (rolling mean, window size 10) over the training episodes of the baseline policy.}
    \label{fig:reward_curves_baseline}
\end{figure}

Fig.~\ref{fig:reward_curves_baseline} reports the reward dynamics of the random baseline across 400 training episodes. In Fig.~\ref{fig:reward_curves_baseline} (a), the per episode reward shows high variability, which is expected under stochastic action selection and stochastic path degradations. Importantly, the reward does not exhibit a consistent improving trend, indicating that the baseline does not adapt its decisions over time. Fig.~\ref{fig:reward_curves_baseline} (b) reinforces this conclusion, since the rolling mean with window size 10 remains relatively flat, providing a stable reference for assessing whether learning yields consistent performance gains.

\begin{figure}[!ht]
    \centering
    \begin{tabular}{@{}c@{}c@{}}
        \includegraphics[width=0.400\columnwidth]{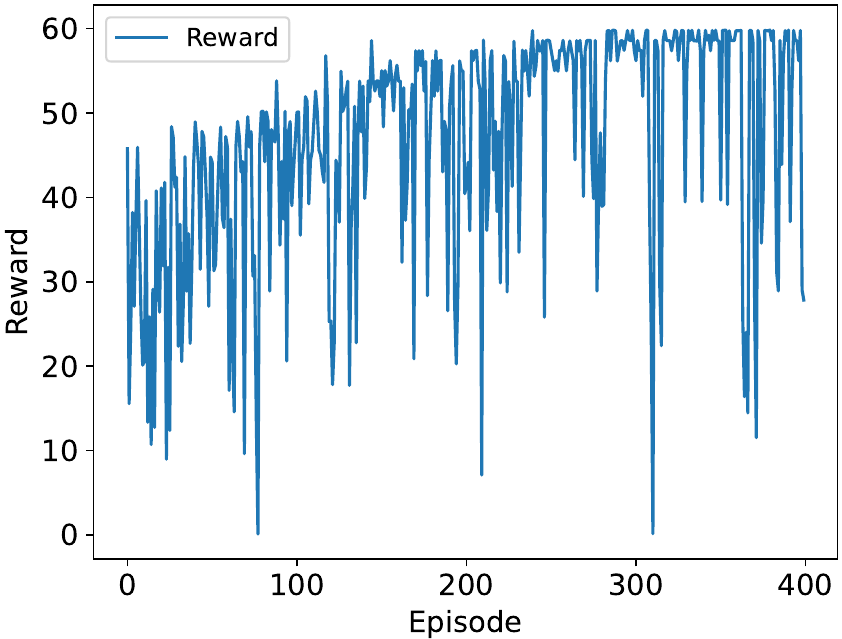} &
        \includegraphics[width=0.400\columnwidth]{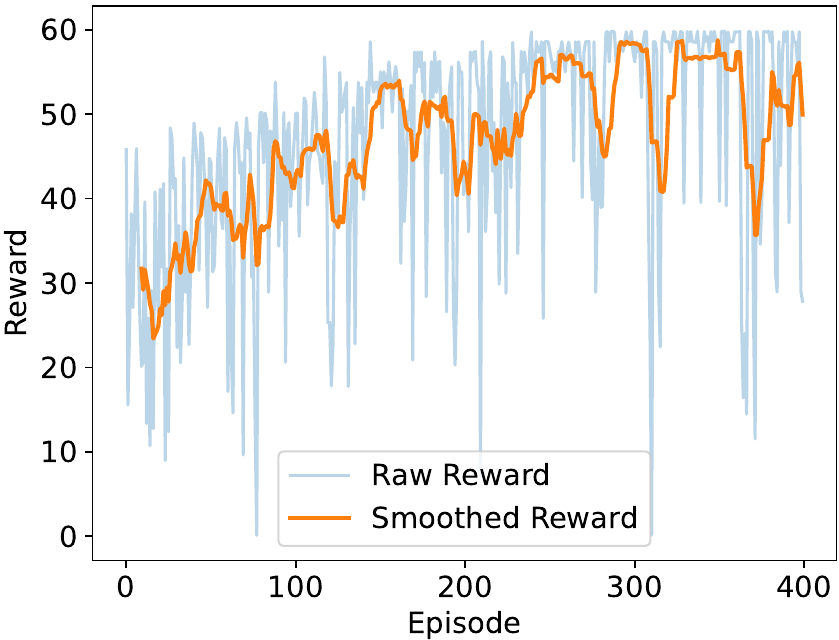} \\
        \small (a) Raw reward per episode. & \small (b) Smoothed reward.  \\
    \end{tabular}
    \caption{Comparison of raw and smoothed cumulative rewards (rolling mean, window size 10) over the training episodes of the \ac{DQN} agent.}
    \label{fig:reward_curves}
\end{figure}

\textbf{\ac{DQN} learns an effective and stable low latency policy under stochastic degradations}. Fig.~\ref{fig:reward_curves} shows the reward evolution of the \ac{DQN} across the same 400 training episodes. The raw reward signal in Fig.~\ref{fig:reward_curves} (a) is initially noisy due to exploration and the inherent variability of the environment. When applying a rolling mean with window size 10 in Fig.~\ref{fig:reward_curves} (b), a clear upward trend emerges, with rewards improving steadily until convergence around episode 250. This behavior indicates that the agent successfully learns a policy that consistently achieves higher reward, which corresponds to lower observed delay, while occasional drops remain plausible due to exploration and transient adverse conditions.

\begin{figure}[!ht]
    \centering
    \begin{tabular}{@{}c@{}c@{}}
        \includegraphics[width=0.450\columnwidth]{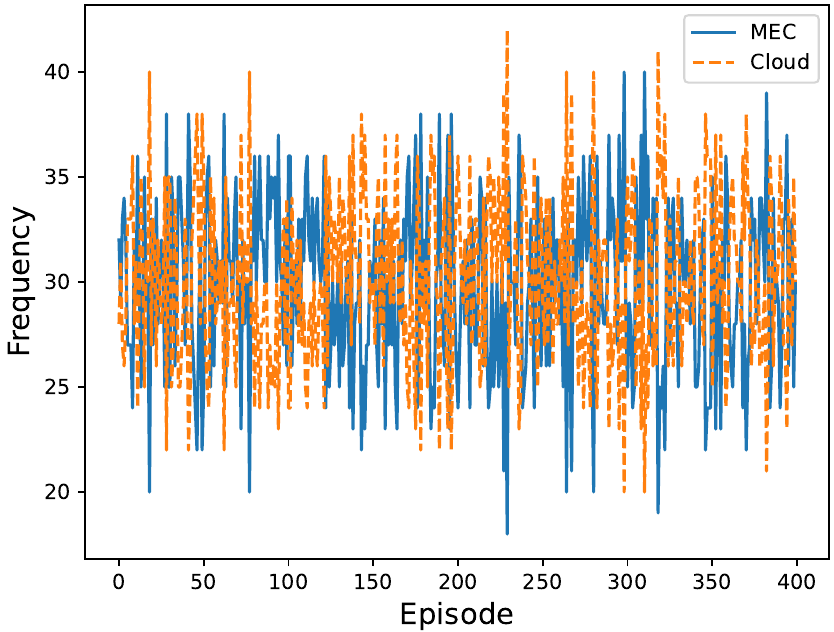} &
        \includegraphics[width=0.450\columnwidth]{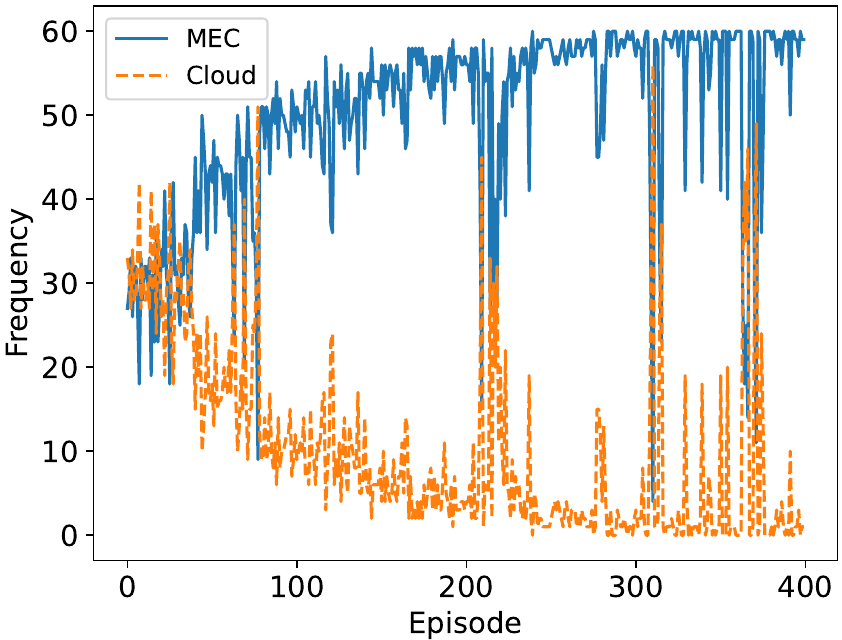} \\
        \small (a) Baseline policy. & \small (b) \ac{DQN} policy. \\
    \end{tabular}
    \caption{Action distribution across all training episodes for both the baseline and the \ac{DQN} policies.}
    \label{fig:action_distribution}
\end{figure}

Fig.~\ref{fig:action_distribution} connects learning dynamics to policy behavior by showing action distributions over the 400 episodes. The baseline in Fig.~\ref{fig:action_distribution} (a) alternates between the \ac{MEC} path and the cloud path without a structured pattern, consistent with random selection. In contrast, Fig.~\ref{fig:action_distribution} (b) shows that the \ac{DQN} increasingly favors the \ac{MEC} interface, which was configured with lower latency and jitter, while progressively reducing reliance on the cloud path. This shift indicates that the agent identifies the more reliable interface and sustains this preference as learning progresses, rather than oscillating between choices.

\begin{figure}[!ht]
    \centering
    \begin{tabular}{@{}c@{}c@{}}
        \includegraphics[width=0.45\columnwidth]{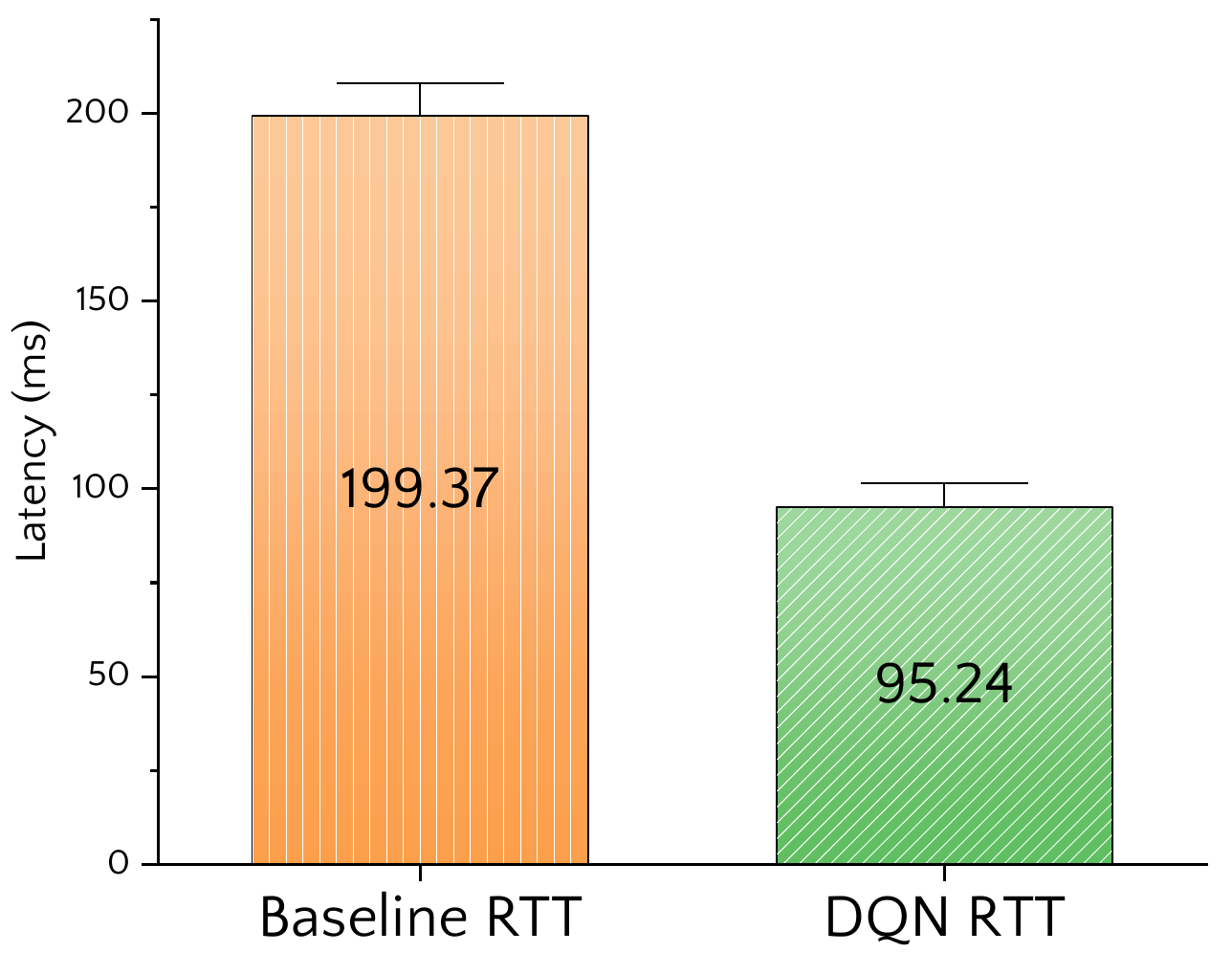} &
        \includegraphics[width=0.45\columnwidth]{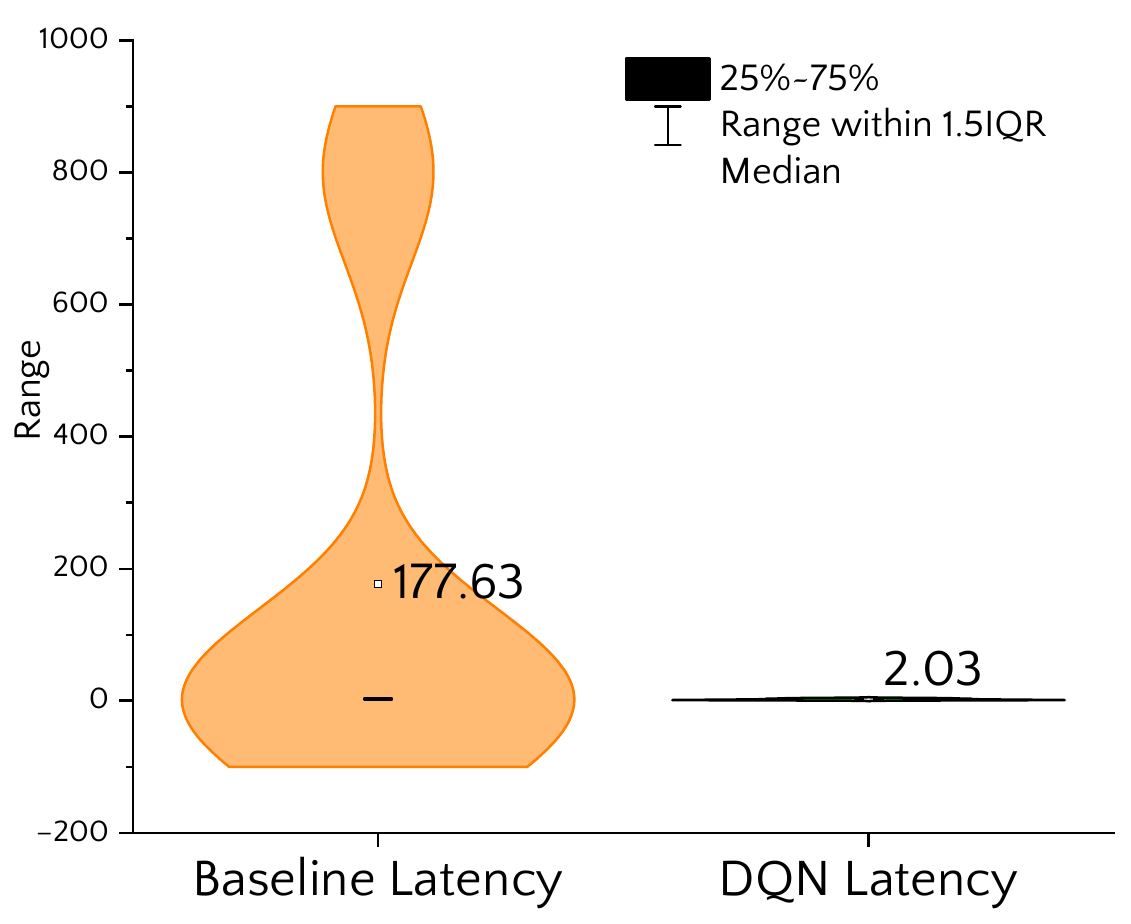} \\
        \small (a) Full episodes. & \small (b) Last 50 episodes. \\
    \end{tabular}
    \caption{Comparison of average round trip time.}
    \label{fig:rtt_comparison}
\end{figure}

Fig.~\ref{fig:rtt_comparison} quantifies the latency impact of these decisions. Over all episodes in Fig.~\ref{fig:rtt_comparison} (a), the \ac{DQN} attains a substantially lower mean round trip time, 95.24 ms, compared to 199.37 ms for the baseline, demonstrating its ability to exploit the lower latency interface. Focusing on the last 50 episodes in Fig.~\ref{fig:rtt_comparison} (b), the contrast becomes stronger in terms of stability, the \ac{DQN} maintains a mean variation of 2.03 ms, whereas the baseline exhibits a volatile range of 177.63 ms. Together, these results show that the learned policy improves both average latency and temporal consistency, while the baseline remains erratic under the same stochastic conditions. These results show that the proposed e\ac{UPF} effectively supports time-sensitive path selection by consistently favoring the most suitable interface under dynamic conditions.

\begin{figure}[htbp]
\centering
  \includegraphics[width=0.55\columnwidth]{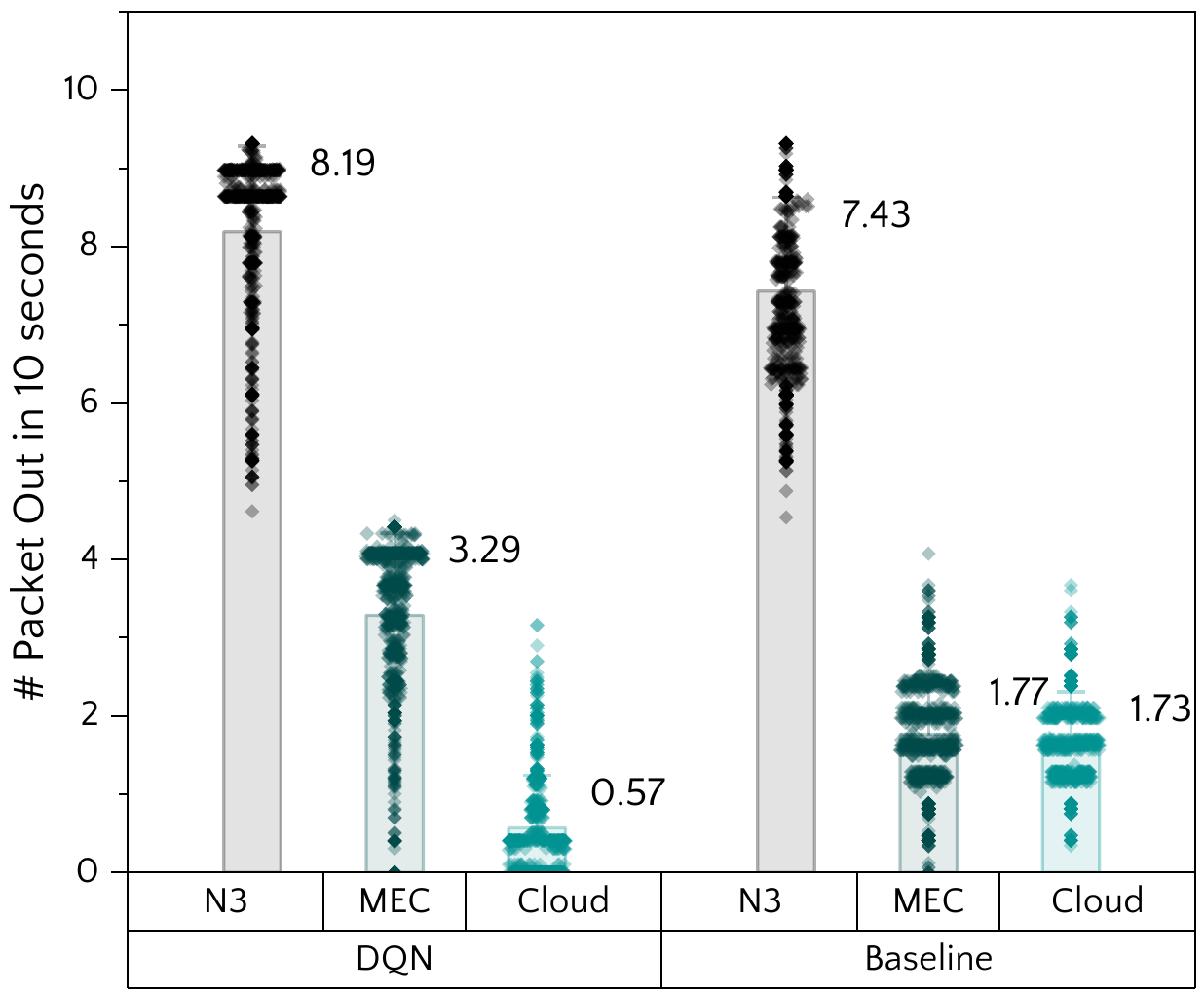}
  \caption{Packet out distribution.}
  \label{fig:packet_out_distribution}
\end{figure}

\textbf{Learned steering decisions materialize as consistent data plane forwarding toward the best path}. Fig.~\ref{fig:packet_out_distribution} presents the distribution of packets forwarded through each interface under both policies. During the experiment, the User Equipment generated Internet Control Message Protocol traffic at one packet per second, and the plot aggregates transmissions over 10 second intervals, averaged across the entire duration. Under the \ac{DQN} policy, more packets are forwarded to \ac{MEC}, with an average of 3.29 per interval versus 0.57 toward the cloud, reflecting sustained preference for the lower latency interface. In contrast, the baseline remains nearly balanced, with 1.77 and 1.73 packets per interval toward \ac{MEC} and cloud respectively, consistent with uniform random decisions. This forwarding behavior matches the latency outcomes, confirming that the agent translates observed delay into actionable interface selection decisions at the data plane.

\begin{table}[htbp]
\centering
\caption{Descriptive statistics for \ac{DQN} and baseline in the last 50 episodes.}
\begin{tabular}{lcc}
\hline
\textbf{Metric} & \textbf{\ac{DQN} (± SD)} & \textbf{Baseline (± SD)} \\
\hline
Average Reward           & 49.40 ± 14.92     & 30.61 ± 9.31      \\
Average Latency [ms]     & 298.46 ± 443.81   & 486.57 ± 398.46   \\
Reward Median            & 58.59             & 33.25             \\
Latency Median [ms]      & 6.95              & 453.86            \\
Reward Range             & [11.53, 59.84]    & [10.41, 44.18]    \\
Latency Range [ms]       & [4.82, 1439.40]   & [5.01, 1290.59]   \\
\hline
\end{tabular}
\label{tab:descriptive_stats_dqn_baseline}
\end{table}

Table~\ref{tab:descriptive_stats_dqn_baseline} summarizes the last 50 episodes, where the reinforcement learning policy is already stabilized. The \ac{DQN} achieves higher rewards, 49.40 $\pm$ 14.92 versus 30.61 $\pm$ 9.31, and lower average latency, 298.46 $\pm$ 443.81 ms versus 486.57 $\pm$ 398.46 ms. The median latency further highlights stability gains, 6.95 ms for the \ac{DQN} versus 453.86 ms for the baseline. The ranges indicate that both policies can still encounter occasional high delay events due to induced degradations, yet the \ac{DQN} concentrates more decisions on the better performing interface, improving overall delay and reducing variability relative to the baseline.

\section{Concluding Remarks}\label{sec:concluding_remarks}

While \ac{B5G} \ac{SLA} management often relies on simulation based admission control, it lacks real time intelligent intervention in the data plane. Our evaluation on the FABRIC testbed demonstrates that the \ac{DQN} based agent achieves stable convergence and significantly reduces latency compared to random baselines. These results indicate that transforming the \ac{UPF} into an intelligent \textit{e}\ac{UPF} effectively enables autonomous slice-aware and time-sensitive path selection in \ac{B5G} edge networks.

Although this study focuses on uplink steering within a specific testbed scenario, it validates the feasibility of non intrusive telemetry for \ac{RL} observations. Future research will investigate alternative algorithms and packet queuing prioritization to enhance efficiency across both uplink and downlink traffic. This research provides a critical foundation for programmable architectures and intelligent network management in the era of low latency mobile services.

\begin{credits}
\subsubsection{\ackname} The authors thank the National Council for Scientific and Technological Development (CNPq) under grant number 421944/2021-8 (call CNPq/ MCTI/ FNDCT 18/2021), FAPEMIG (Grant \#APQ00923-24), FAPESP MCTIC/CGI Research project 2018/23097-3 - SFI2 - Slicing Future Internet Infrastructures. FCT has also supported this work – Fundação para a Ciência e Tecnologia within the R\&D Unit Project Scope UID/00319/Centro ALGORITMI (ALGORITMI/UM).

\subsubsection{\discintname}
The authors have no competing interests to declare that are relevant to the content of this article.
\end{credits}

\bibliographystyle{splncs04}
\bibliography{references}

\end{document}